\documentclass{iaus}
\usepackage{graphicx}

\title[CNO in the bulge] 
{CNO abundances in the Galactic bulge}

\author[N. A. E. Ryde]   
{Nils Ryde$^1$
}

\affiliation{$^1$Lund  Observatory, Box 43, SE-221 00 Lund, Sweden\\ email: {\tt ryde@astro.lu.se} }

\pubyear{2009}
\volume{265}  
\jname{Chemical Abundances in the Universe: Connecting First Stars
to Planets}
\editors{K. Cunha, M. Spite \& B. Barbuy , eds.}
\begin{document}

\maketitle

\begin{abstract}
The carbon, nitrogen, and oxygen abundances and  trends in the bulge are discussed in the context of  our recent analysis of these elements in an on-going project based on near-IR spectra (Ryde et al. 2009). We obtained these  using the CRIRES spectrometer on the VLT. The formation and evolution of the Milky Way bulge can be constrained by studying elemental abundances of
bulge stars. Due to the large and variable visual extinction in the line-of-sight towards the bulge, an analysis in the near-IR is preferred.
\keywords{stars: abundances, Galaxy: bulge, infrared: stars}
\end{abstract}


It is strange, but we actually do not know the origin and the classification of the Milky Way bulge. There are two main avenues for the formation of bulges, one leading the the pseudobulges and one to the classical ones (see, e.g. Kormendy \& Kennicutt, 2004). The former are formed through secular, dynamical evolution of the disks over time, driven by the development of a bar. One would then expect an age spread of its stars.
The dynamics and morphology of the Milky Way bulge places it as pseudobulge.
Its shapes and structures (boxy, peanut-shaped bar) indicates this.
Indeed, Binney (2009) writes that `...there is every indication that our bulge is a pseudobulge. For this reason alone it would 
be connected to the disc by history.' The second avenue leads to the formation of the classical bulges, which are formed by merger-driven star-bursts during a short phase. The stars of the Milky Way bulge are old, the bulk being $\sim11$ Gyrs (Ortolani et al 1995, Clarksson et al. 2009), even though there is  a younger population at the centre (see e.g. Figer et al. (2004)).  Furthermore,  the stars are metal-rich and are $\alpha$-element enhanced. This indicates that the bulge underwent rapid chemical enrichment as the result of starbursts, as in the case of a classical bulge.  Thus the formation of the Milky Way bulge is clearly not well understood and its classification inconclusive.  Classical bulges are most often found in early-type spirals (Sa and Sb) whereas the pseudobulge are found in late-type spirals (Sc and Sd). The Milky Way is classified as a Sbc and is morphologically on the borderline. It is therefore maybe not a surprise that the origin of it is under debate. Maybe the bulge formed from a secular, but fast evolution of the 
early disk, resulting in an early formation (Genzel et al. 2008).


The different formation scenarios can be
constrained by abundance surveys. From stellar populations and abundance analyses it can be investigated
which process dominated the star formation:
whether stars formed rapidly a long time ago during hierarchical clustering of galaxies (as a classical bulge), or through secular evolution (such as defines a pseudo-bulge). The $\alpha$-element composition (e.g. O, Mg, Ca, etc.) relative to iron as a function of the metallicity, [Fe/H], can  infer star-formation rates (SFR) and initial-mass functions (IMF). A shallower IMF will increase the number $\alpha$-element producing stars thus leading to higher [$\alpha$/Fe] values. A faster enrichment due to a high star-formation rate will keep the over-abundance of the $\alpha$ elements at a high value also at higher metallicity. Different populations may show different behaviours.
However, the chemical properties and evolution of the MW bulge are poorly constrained, mainly due to the large distance and large and variable visual extinction toward the bulge. We, therefore need more stellar abundance studies of many more bulge stars of various bulge fields, especially in the near-IR, to constrain bulge models.

The Bulge
is more accessible in the IR than in the optical for multiple reasons
(Ryde et al. 2005). 
The most important is of course the smaller interstellar extinction in
the IR ($\mathrm{A_{K} \sim 0.1 \times A_V}$; Cardelli et al. (1989)). 
The near-IR is also preferred for analysis of abundances, due to
the fact that the absorption spectra are less crowded with lines, that fewer lines are blended, and that it is
easier to find portions of the spectrum which can be used to define a continuum compared to wavelength regions in the optical spectral window. Furthermore, in the Rayleigh-Jeans regime, the
intensity is less sensitive to temperature variations. This means that
the effects of, for example, effective-temperature uncertainties or
surface inhomogeneities on line strengths should be smaller in the
IR. Note, however, that 3D effects on the atmosphere can have an impact though the sensitivness of the molecular equilibria to temperature.
A general drawback of a spectral analysis in the near-IR is that 
existing HR spectrometers are still much less effective than optical ones, one
of the main reasons being the lack of cross-dispersion.
Also, determining the stellar parameters based only on near-IR spectra is difficult. 
The main advantage of a high spectral resolution (HR, $R>50,000$) is the smaller line-blending, reduction of background noise, and the fact that telluric lines can be taken care of more easily.

Especially, the CNO abundances are of interest and infrared spectroscopy at high spectral resolution can provide the most secure measurements of the CNO elements. Oxygen provides information on the SFR and IMF through O/Fe and C/O ratios. The C+N abundances give an indication of the evolutionary stage of the red giants investigated, and 
N is needed to estimate the ÔprimordialÕ C abundance of the red giants.
The C abundance can give information on its origin in the bulge. To get the CNO abundances, normally one needs to measure them simultaneously due to the molecular equilibria.
 Recently,  a few studies of elemental abundances of bulge stars using near-IR spectra at high resolution have been done, see for instance Ryde et al. (2009), Mel\'endez et al. (2008), Cunha et al. (2007), Cunha \& Smith (2006), and Mel\'endez et al. (2003).

We\footnote{B. Gustafsson (PI), B. Edvardsson, J. Melendez,
A. Alves Brito, M. Asplund, B. Barbuy, V.  Hill, H.-U. K\"aufl, D. Minniti, S. Ortolani, A. Renzini, N. Ryde, and M. Zoccali}  are engaged in an on-going VLT project (080.D-0675),  in which we are analysing near-IR spectra recorded   at a spectral resolution of  $R=60,000$ with the CRIRES spectrometer (Moorwood 2005; K\"aufl et al. 2006)  
on the {\it Very Large Telescope, VLT}, see also Ryde et al. (2007, 2009). 
We have measured the
abundances of Fe,  C, N, and O in addition to the $\alpha$ elements Si,
S, and Ti  for a well-chosen
sample of 11 stars, sampling different stages of
the chemical enrichment history and different parts of the bulge. The stars are chosen from the optical work by Lecureur et al. (2007) and Zoccali et al. (2006, 2008). The stars lie in three fields at $b=-3$ (NGC 6553 field), $-4$ (BW), and $-6$ and have $H$ magnitudes between 10.3 and 12.0. The spectra are analysed with tailored MARCS models and fully consistent synthetic spectra. The temperatures lie between $3900-4300$~K, $-1.25<$[Fe/H]$<0.0$, $1.0<\log g < 2.2$.  The quality of the spectra are very high and numerous molecular lines of CO, CN, and OH can be identified, see Ryde et al. (2009). We have also identified 10 Si, 5 Ti, 4 S, 5 Ni, one Cr,
and many Fe lines in the spectra.

We find a high [O/Fe] vs. [Fe/H] trend of 0.4 up to metallicities of [Fe/H$=-0.3$ whereafter the [O/Fe] declines. The data suggest that the Milky Way bulge experienced a rapid and early star-formation history compared to the thin disk. Our data agrees well with other determinations of the oxygen abundances in the bulge both from near-IR lines (Melendez et al.(2008), Cunha \& Smith (2006), and Rich \& Origlia (2005)) and from optical lines (Fulbright et al. (2007) and  Lecureur et al. (2007) and Zoccali et al. (2006)). Our stars are a subsample of the latter. Our and Melendez et al.Õs data confirm the result of Zoccali et al.  When comparing with thick and thin disks giants from Melendez et al. (2008) we find that the thick disk and bulge trend are similar. Thus, both the bulge and thick disk must have formed rapidly (Melendez et al. 2008). A key point is that one can reduce the systematic uncertainties by homogeneously comparing disk giants with bulge giants on the same scale (Melendez et al. 2008).

\begin{figure}[b]
\begin{center}
 \includegraphics[width=4in,angle=90]{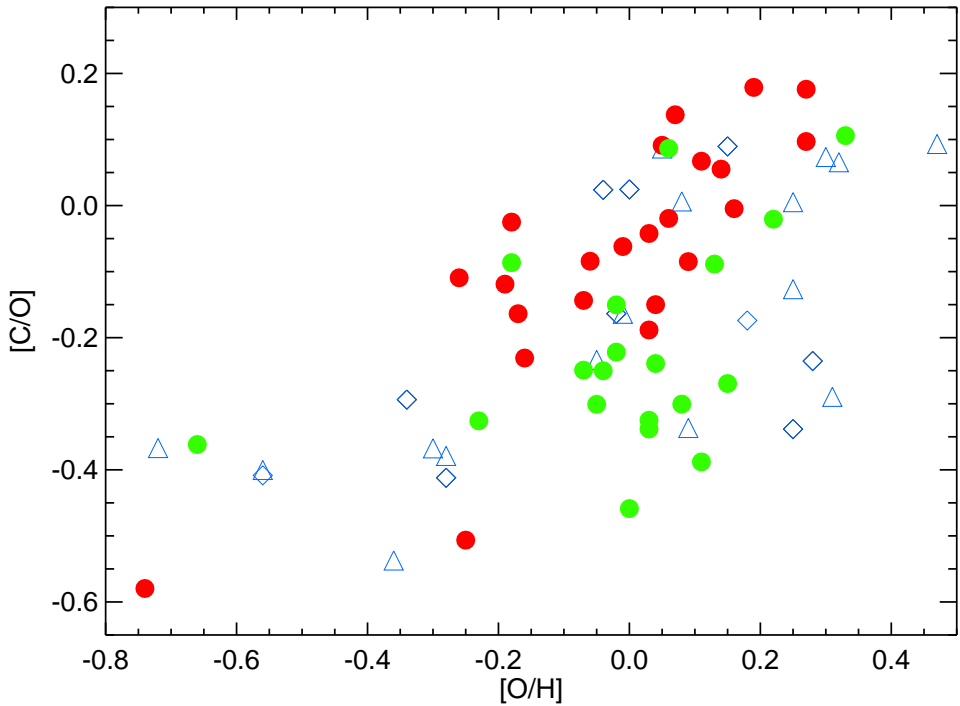} 
 \caption{[C/O] vs. [O/H] for our bulge giants (dimonds) and those from Melendez et al.  2008 (triangles) shown with blue symbols. The thin and thick disk giants from Melendez et al. (2008) are shown with red and green circles, respectively.  The thick disk giants and the bulge giants show a similar behavior. }
   \label{fig2}
\end{center}
\end{figure}

Recently,  Cescutti et al. (2009)  showed in a plot of  primordial [C/O] vs. [O/H] for thin and thick  disk dwarfs and turnoff stars and  bulge giants a  similar similarity between the bulge and thick disk relationships. The primordial C abundance is estimated by calculating C+N $-$ Fe$\times$(N/Fe)$_\odot$. In Figure \ref{fig2} we show our relationship in the same [C/O] vs. [O/H] diagram.  The difference is that we compare with the Melendez et al. (2008) thin and thick disk giants which are on the same scale. We also find a similar relationship and an upturn, but our data show a larger scatter.

To conclude, we have analysed near-IR spectra from the CRIRES spectrometer on the VLT with tailored MARCS model atmospheres and consistent synthetic spectra. We find that the stellar parameters are important for the derivation of the CNO abundances from molecular lines.
We find [O/Fe] vs. [Fe/H] values enhanced up to metallicities of [Fe/H]$=-0.3$, implying a rapid and early star formation in the bulge, as for a classical bulge.  The [C+N/Fe] is nearly constant with metallicity implying that the oxygen abundance we measure is the primordial, unprocessed values, and that our target stars are indeed first ascent giants or clump giants. We show that the [O/Fe] vs. [Fe/H] trends in the literature all corroborate each other within uncertainties.
The bulge and thick disk seem similar in [O/Fe] vs. [Fe/H] and [C/O] vs. [O/H], implying that both formed rapidly. Such a similarity could suggest that the bulge has a pseudobulge origin. Note, that the metallicity distributions are not the same.  
We do not see an increased [C/Fe] but corroborate an increase in the [C/O] vs. [O/H] plot for [O/H]$>-0.1$.

Stellar surface abundances in Bulge stars, especially those of the C, N, and O elements, can extensively be studied in the near-IR, due to lower extinction. It will be very important to extend the analysis to 
other regions of the Galactic bulge, such as in the galactic plane, in order to get a proper handle on its formation and evolution. In these regions the optical extinction is large which only permits observations in the near-IR. 
Near-IR, high-spectral-resolution spectroscopy offers a promising methodology to study the whole bulge to give clues to its formation and evolution. 



\end{document}